# Single-layer Circular SIW Filtenna With Beam Scanning Capability for 5G Millimeter Wave Communication Applications

Jiawang Li, Hanieh Aliakbari, *Senior Member, IEEE*

**Abstract**—In this communication, two novel low-cost single-layer filtering antennas (filtennas) are proposed for millimeter wave (mmWave) applications. The proposed filtennas consists of a compact circular substrate integrated waveguide (SIW) cavity, a metal post close to the center of the cavity for power feeding, a metal post in the center for modes controlling, and a slot for radiating power. In the passband, the fundamental $TM_{010}$ mode and the $TM_{110}$ mode in the circular SIW cavity are excited by the feeding post. In addition, thanks to the high-pass characteristics of the cavity, it exhibits more than 20 dB suppression in the lower frequency band. There are three radiation nulls in Filtenna I and one radiation null in Filtenna II in the upper band which increase the suppression level as high as 18 dB. As a proof of concept, the proposed filtennas are fabricated and measured. It is shown that the Filtenna I can achieve simulated and measured -10 dB impedance fractional bandwidth (FBW) of 7.1% (27.14 − 29.13 GHz) and 8.6% (27.62 − 30.11 GHz), respectively. While filtenna II can achieve simulated and measured -10 dB FBW of 7.4% (27.86 − 29.99 GHz) and 10.1% (28.11 − 31.09 GHz), respectively. The filtennas features stable radiation patterns with an average gain of 5.0 dBi. The lower and upper sideband suppression levels for both filtennas exceed 18 dB. These filtennas are good candidates for 5G mmWave applications, as they simultaneously provide beam scanning and filtering capability with a low cost, and single layer structure.

## I. INTRODUCTION

**M**odern wireless communication is developing so quickly that there is a growing need for compact systems with high efficiency, and stability [1]. Considering this, communication systems that integrate several components have gained popularity because of their appealing qualities of small size and great efficiency. Meanwhile, since the antenna and bandpass filter are two essential components in the radio frequency (RF) ends, the filtering antenna (filtenna), which processes both filtering and radiating functions, has been studied extensively [2]. According to the current 3GPP standards [3], 5G millimeter wave (mmWave) can support large bandwidth of 400MHz/800MHz. Therefore, designing a low cost filtenna with stable gain, good filtering response within a frequency bandwidth of more than 800 MHz, is desirable for the 5G mmWave applications. Recently, several types of filtennas have been reported [4]-[22], which are based on patch antennas [4]-[6], cavity antennas [7]-[19], and dielectric resonator antennas [20]-[22]. While filtenna designs at lower microwave frequencies have been widely reported, an increasing number of studies have also explored filtennas operating at mmWave frequencies [4], [8], [9], [12]-[19]. It has been shown in [4], [12], [18] that an impedance bandwidth of more than 20% and out-of-band suppression of more than 14 dB can be achieved, but at the cost of using complex and multilayer structures. Furthermore, the filtennas proposed in [12] are bulky with more than half a wavelength. In [18] cavity cascade structure has been proposed which results in a gain of only 4.5 dBi. Out-of-band suppression can exceed 30dB for use in N258 [7], but this method introduces radiation nulls by adding vias to the SIW transmission line in advance and introducing more filter cavities, which greatly increases the antenna area. More than 35 dB stopband suppression is obtained in [15], but its operating bandwidth is only 500 MHz. In [17] and [19] a high gain of more than 10 dBi is achieved using a simple two PCB layers. However, their large component area prevents the antenna from having beam scanning capabilities, which limits the application of the antenna. Therefore, designing a filtenna with a small area, low cost, and beam scanning function is challenging for 5G mmWave applications.

In this communication, two novel single-layer mmWave circular SIW filtenna are proposed for the first time. The filtennas uses cavity modes only to achieve bandwidth expansion. By introducing a curve slot at a cavity, dual high-order modes can be merged, significantly enhancing the bandwidth of the passband. The main innovation points are as follows:

1- Low-cost single-layer PCBs are used to achieve dual-mode operation using cavity modes (i.e.$TM_{010}$ and $TM_{110}$).

2- Compared with multi-mode filtennas, this filtenna not only meets the 800 MHz bandwidth requirements in the current mmWave 5G standard but also has good out-of-band filtering characteristics.

3- The filtenna is small, supports a scanning range of ±40°, and different beams have similar frequency responses.

The proposed filtennas were fabricated and experimentally evaluated. Due to their simple structure, low cost, and excellent filtering performance, they are well-suited for mmWave applications.

## II. FILTENNA CONFIGURATION AND OPERATING MECHANISM

### A. Modes Merging

To demonstrate the operating mechanism of the proposed Filtenna, the modes inside the circular metal cavity have been investigated in this section. A circular metal cavity in the simulations is based on PCB with the thickness of 1.0 mm and it is made of F4B substrate ($\varepsilon_r = 3.55$, $\tan\delta = 0.004$). The resonant frequency of the full-mode SIW (FMSIW) circular cavity is given as follows [9],

$$f_{mn0} = \frac{c}{2\pi\sqrt{\mu_r \varepsilon_r}} \frac{P_{nm}}{R_{eff}} \qquad (1)$$

where $f_{mn0}$ is the resonant frequency of the corresponding mode, $c$ is the velocity of the light in the vacuum, $\varepsilon_r$ and $\mu_r$ are the relative permittivity and permeability of the substrate, respectively, $P_{nm}$ stands for the $m$th root of the $n$th-order Bessel function of the first kind and $R_{eff}$ is the effective radius of the conventional circular cavity given by [9]:

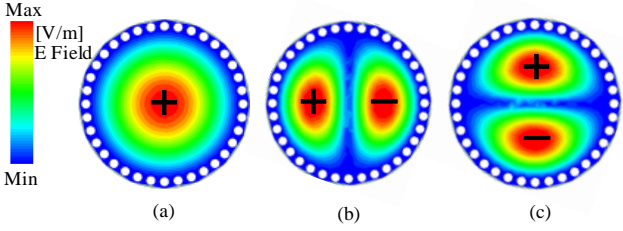

Fig. 1. E-field distribution of FMSIW cavity at (a) TM$_{010}$ mode at 18.2 GHz, (b) TM$_{110}$ mode at 29.0 GHz, and (c) TM$_{110}$ mode at 29.0 GHz.

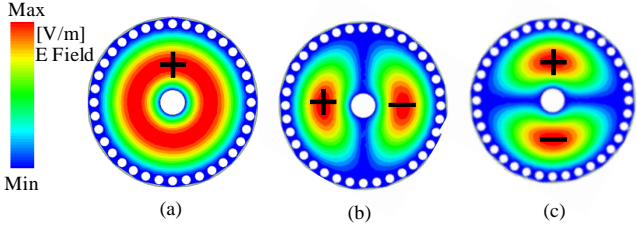

Fig. 2. E-field distribution of FMSIW cavity after loading via in the center at (a) TM$_{010}$ mode at 27.4 GHz, (b) TM$_{110}$ mode at 31.1 GHz, and (c) TM$_{110}$ mode at 31.1 GHz. (A metal hole with a diameter of 1.1 mm is inserted.)

$$R_{eff} = r - 1.08\left(\frac{d^2}{s}\right) + 0.1\left(\frac{d^2}{r}\right) \tag{2}$$

where $d$ and $s$ are the diameter and spacing between the adjacent vias, respectively. Fig. 1 shows the simulated E-field modal distribution of the FMSIW cavity. The cavity radius is 3.5 mm. High Frequency Structure Simulator (HFSS) 2020 R1 is used to get the results. Since the aim is to support a beam scanning function, it is necessary to select the fundamental mode and the higher-order modes which have closer resonant frequencies to the fundamental mode. For a circular resonant cavity, the first three resonant modes are TM$_{010}$, TM$_{110}$ and its degenerate mode. Their corresponding resonant frequencies are 18.2 GHz, and 29 GHz respectively. The simulated frequency ratio of TM$_{110}$ and TM$_{010}$ is 1.5934, which is very close to the theoretical value of 1.59 from (1). However, the frequency difference between these two modes is large, making it difficult to combine the two modes. One way is to introduce additional modifications in the cavity which has a small impact on the resonant frequency of one of the modes, while it changes the resonant frequency of the other mode significantly, so that the frequencies of the two resonant modes are close to each other. Therefore, a metal hole is introduced in the center of the cavity, which has more impact on the resonant frequency of the TM$_{010}$ compared to the TM$_{110}$ and its degenerate mode. That is because TM$_{010}$ has high E-field at the center (see Fig. 1(a)) while TM$_{110}$ and its degenerate mode have low E-field at the same position (see Fig. 1 (b) and (c)).

As shown in Fig. 2, the cavity has been changed to an SIW coaxial structure, when a metal hole is loaded in the center of the cavity. According to the boundary condition:

$$E_{z,\rho=a} = E_{z,\rho=b} = 0 \tag{3}$$

where $a$ and $b$ are the radius of the cavity and and radius of the central metal hole, respectively. The transcendent equation of the TM wave can be obtained as [23]:

$$J_n(k_c a) N_n(k_c b) = J_n(k_c b) N_n(k_c a) \tag{4}$$

$k_c$ is the cut-off wave number. $J_n(\cdot)$ and $N_n(\cdot)$ are the Bessel and Neuman functions, respectively. From (3), the electric field expression of the TM wave in the SIW cavity after loading the metal hole can be obtained as:

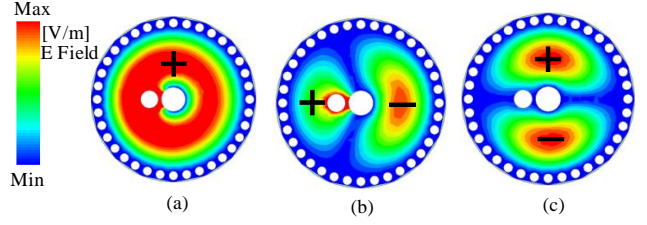

Fig. 3. E-field distribution of FMSIW cavity at (a) TM$_{010}$: 27.1 GHz, (b) TM$_{110}$ mode: 30 GHz, and (c) TM$_{110}$ mode: 31.3 GHz.

$$E(r,\phi,z) = E_0\left[N_n(k_c a)J_n(k_c r) - J_n(k_c a)N_n(k_c r)\right]\sin(m\phi)e^{-j\beta z} \tag{5}$$

$E_0$ is the peak value of the electric field and $\beta = 2\pi/\lambda$ is the phase constant, $\lambda$ is the corresponding wavelength. Therefore, the resonant frequency of the TM$_{010}$ mode in SIW coaxial structure is approximately as:

$$f_{TM_{010}} = \frac{P_{01}c}{2\pi(a-b-\Delta R)\sqrt{\mu_r\varepsilon_r}} \tag{6}$$

where $\triangle R \approx 2(r-R_{eff})$. The calculated resonant frequency of TM$_{010}$ mode is 27.193 GHz, which is close to the simulated results (i.e. 27.4 GHz in Fig. 2 (a)). Therefore, the resonant frequency of the TM$_{010}$ mode can be changed by changing the radius of the inserted via hole.

On the other hand, using via with smaller radius can be considered as a perturbation to the TM$_{110}$ mode. According to the cavity perturbation theory, there is the following expression [23]

$$\frac{f_{TM_{110}} - f_{TM_{110}}^0}{f_{TM_{110}}} = -\frac{\int\limits_{\Delta V}\left(u_0\left|H_{110}\right|^2 - \varepsilon_0\left|E_{110}\right|^2\right)dV}{\int\limits_{V}\left(\varepsilon_0\left|E_{110}\right|^2 + u_0\left|H_{110}\right|^2\right)dV} \tag{5}$$

$$= -\frac{\int\limits_{\Delta V}\left(u_0\left|H_{110}\right|^2 - \varepsilon_0\left|E_{110}\right|^2\right)dV}{2\int\limits_{V}\varepsilon_0\left|E_{110}\right|^2 dV}$$

where $f_{TM_{110}}$ and $f_{TM_{110}}^0$ are the resonant frequencies after and before loading perturbation metal, respectively. $\Delta V$ is the volume of the inserted via hole and $V$ is the volume of the circular cavity. $u_0$ and $\varepsilon_0$ is the permeability and permittivity in free space. In a circular SIW cavity, the expressions of the TM$_{110}$ mode for both electric field and magnetic field can be obtained as [23]:

$$\begin{cases} E_r = E_\phi = H_z = 0 \\ E_z = E_{110}J_1(k_{11}\rho)\cos(\phi) \\ H_r = -\left(\frac{iw\varepsilon_0}{k_{11}^2\rho}\right)E_{110}J_1(k_{11}\rho)\sin(\phi) \\ H_\phi = -\left(\frac{iw\varepsilon_0}{k_{11}}\right)E_{110}J_1(k_{11}\rho)\cos(\phi) \end{cases} \tag{6}$$

where $k_{11} = \omega\sqrt{u_0\varepsilon_0}$ and $J_n(\cdot)$ is the $n$th-order Bessel function. Therefore, through three-dimensional integration, we can obtain:

$$\int\limits_{V}\varepsilon_0\left|E_{110}\right|^2 dV = \varepsilon_0\iiint\limits_{V}\left|E_{110}\right|^2 dV = \frac{\pi\varepsilon_0 hE_{110}^2 a^2}{2}J_0^2(P_{11}) \tag{7}$$

$$\int\limits_{\Delta V}\left(\mu_0\left|H_{110}\right|^2 - \varepsilon_0\left|E_{110}\right|^2\right)dV = \mu_0\iiint\limits_{\Delta V}\left(\left|H_r\right|^2 + \left|H_\phi\right|^2\right)dV$$

$$= \frac{\pi\varepsilon_0 hE_{110}^2 a^2}{2P_{11}^2}\left[\left(P_{11}Q\right)^2 J_0^2(P_{11}Q) + \left(\left(P_{11}Q\right)^2 - 2\right)J_1^2(P_{11}Q)\right] \tag{8}$$

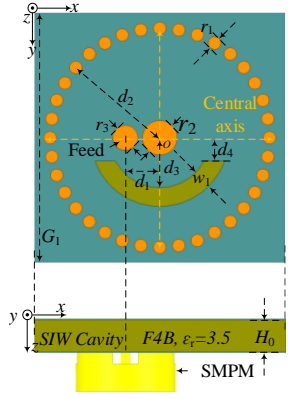

Fig. 4. Topology of the proposed Filtenna I. The dimensions are ($G_1 = 8$, $d_1 = 1.1$, $d_2 = 3.5$, $d_3 = 1.6$, $d_4 = 1.74$, $r_1 = 0.2$, $r_2 = 0.55$, $r_3 = 0.4$, $w_1 = 0.6$, $H_0 = 1.0$, unit: mm.).

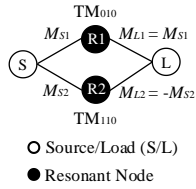

Fig. 5. Coupling topology of proposed antenna.

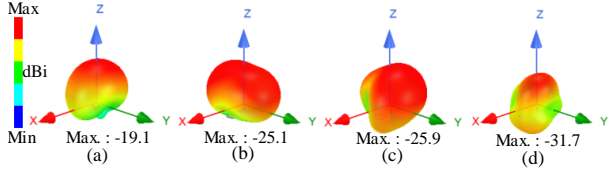

Fig. 6. 3D radiation patterns at the stopband. (a) 20 GHz, (b) 30.3 GHz, (c) 31.6 GHz, (d) 33.2 GHz.

where $Q = (b + \Delta) / (a - \Delta)$ and $\Delta = r - R_{eff}$. In this step, we use the transformations from [24]:

$$\int x J_1^2(x)\, dx = \frac{x}{2}\left[x J_0^2(x) + x J_1^2(x) - 2 J_0(x) J_1(x)\right] \quad (9)$$

$$\int x J_0^2(x)\, dx = \frac{x^2}{2}\left[J_0^2(x) + J_1^2(x)\right] \quad (10)$$

$$\int x J_2^2(x)\, dx = \frac{x^2}{2} J_0^2(x) + \frac{x^2 - 4}{2} J_1^2(x) \quad (11)$$

$$\int_0^a x J_0(x) J_2(x)\, dx = 1 - \frac{1}{2}a^2 J_1^2(a) - \frac{1}{2}(2 + a^2) J_0^2(a) \quad (12)$$

$$J_1'(x) = \frac{1}{2}\left[J_0(x) - J_2(x)\right] \quad (13)$$

Moreover, since the metallic post is placed at the electric field null of the $TM_{110}$ mode, the contribution of the electric field can be neglected in the calculation. Then we can rewrite (5) as:

$$\frac{f_{TM_{110}} - f_{TM_{110}}^0}{f_{TM_{110}}} \approx \frac{(P_{11}Q)^2 J_0^2(P_{11}Q) + \left((P_{11}Q)^2 - 2\right) J_1^2(P_{11}Q)}{2 P_{11}^2 J_0^2(P_{11})} \quad (14)$$

The resulting shifted frequency is calculated to be 31.52 GHz, with a relative error of 1.35%. Suppose the goal bandwidth is $x\%$, by using this method we have:

$$x\% = 2\left(f_{TM_{110}} - f_{TM_{010}}\right) / \left(f_{TM_{110}} + f_{TM_{010}}\right) \quad (15)$$

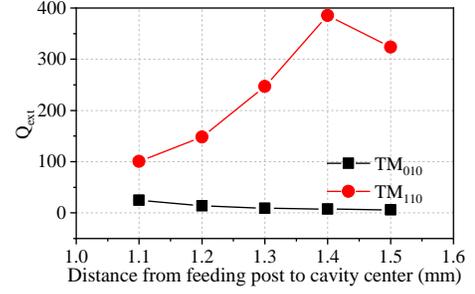

Fig. 7. Extracted $Q_{ext}$ values of the $TM_{010}$ and $TM_{110}$ modes of Filtenna I.

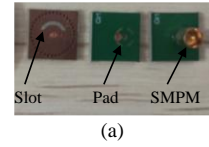

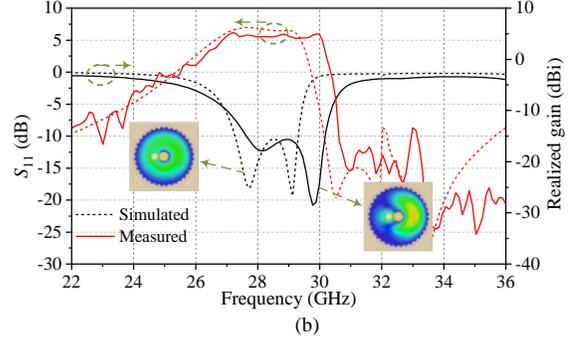

Fig. 8. (a) Photograph of the proposed Filtenna I. (b) Simulated and measured reflection coefficients and realized gains of the proposed Filtenna I.

Then a metal hole with a suitable radius $b$ can be selected to make the resonant frequencies of $TM_{010}$ mode and $TM_{110}$ mode closer.

### B. Antenna Feeding Position Analysis

It is known from the perturbation theory that when a metal hole is introduced at a position with a strong electric field, the resonant frequency will be reduced. Therefore, this property is used to introduce the feed position to achieve a reduction in the resonant frequency of the $TM_{110}$ mode. At the same time, it has a low impact on the resonant frequency of its degenerate mode. It can be seen in Fig. 3 that the introduction of the feed point position has little impact on the resonant frequency of the $TM_{010}$ mode, which is decreased by only 0.3 GHz. However, due to the E-field being strong in the $TM_{110}$ mode for the location where the metal hole is introduced, the impact on resonant frequency is greater, which has been reduced by 1 GHz. The impact on the degenerate mode of $TM_{110}$ mode (Fig. 3(c)) is also very small, and the frequency shifts to high frequency by 0.3 GHz. Therefore, the introduction of the feed position facilitates the proximity between the two modes.

### C. Coupling Topology and Coupling Matrix of Filtenna I

The proposed Filtenna I is depicted in Fig. 4 and is composed of a single-layer circular SIW cavity. The circular SIW cavity is formed by a circle of posts and a half-moon-shaped slot is etched in the top layer. The half-moon-shaped gap effectively facilitates a reasonable excitation of the electric field strength of the desired modes. There is a larger radius post in the center of the cavity for mode controlling. The post offset of the center is used for feeding power. A SMPM connector is used to connect with the feeding post. The antenna is simulated in the High-Frequency Structure Simulator (HFSS) 2020 R1.

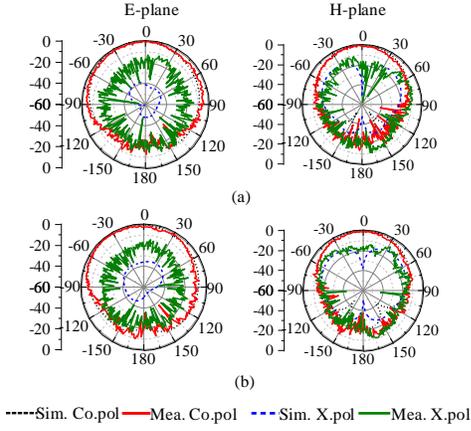

Fig. 9. Simulated and measured normalized radiation patterns of the proposed Filtenna I, (a) 27.6 GHz, (b) 28.9 GHz.

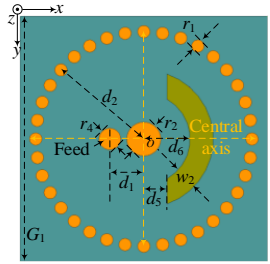

Fig. 10. Topology of the proposed Filtenna II. The dimensions are ($G_1$ = 8, $d_1$ = 1.1, $d_2$ = 3.5, $d_5$ =0.75, $d_6$ = 1.5, $r_1$ = 0.2, $r_2$ = 0.55, $r_4$ = 0.35, $w_2$ = 0.6, unit: mm.).

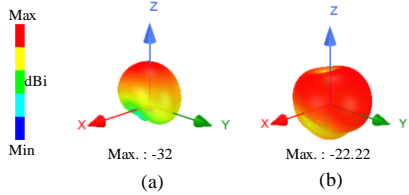

Fig. 11. 3D radiation patterns at the stopband. (a) 20 GHz, (b) 32.8 GHz.

The coupling topology of the proposed Filtenna I is given in Fig. 5. S and L represent the source and load, respectively. And here the load is the radiated slot. R1 and R2 represent two SIW cavity modes, e.g., the $TM_{010}$ mode and $TM_{110}$ mode. In the passband, there are two coupling paths where power can be transmitted from source to load. The first one is S - R1 - L, and another one is S - R2 - L. In this case, $TM_{010}$ mode and $TM_{110}$ mode can be regarded as two coupling resonators according to the coupled resonator theory. The external quality factor $Q_{ext}$ can be expressed as [25]:

$$Q_{ext} = \frac{2\pi\tau f_0}{4} \tag{16}$$

where $f_0$ is the center frequency of the passband. $\tau$ is the group delay of the $S_{11}$ at $f_0$, which is mainly controlled by the feeding position of the post and can be adjusted by changing the distance from the feeding post to the center of the cavity. A reference coupling matrix is given as matrix (4) according to the bandpass filter design theory [25].

$$M = \begin{matrix} & S & 1 & 2 & L \\ S \\ 1 \\ 2 \\ L \end{matrix} \begin{bmatrix} 0 & 0.95 & 0.47 & 0 \\ 0.95 & 0.32 & 0 & 0.95 \\ 0.47 & 0 & -2.27 & -0.47 \\ 0 & 0.95 & -0.47 & 0 \end{bmatrix} \tag{17}$$

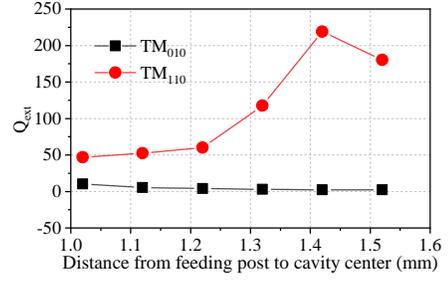

Fig. 12. Extracted $Q_{ext}$ values of the $TM_{010}$ and $TM_{110}$ modes of Filtenna II.

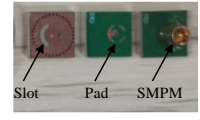

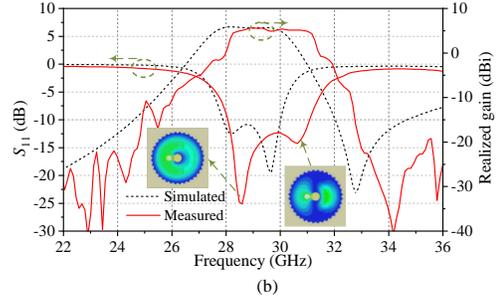

Fig. 13. (a) Photograph of the proposed Filtenna II. (b) Simulated and measured reflection coefficients and realized gains of the proposed Filtenna II.

The target design FBW is 7.1% from 27 to 29 GHz. The return loss within the passband is 12 dB. The calculated $Q_{es1}$ and $Q_{es2}$ are 24 and 98, respectively. These correspond to the external quality factors of the two modes $TM_{010}$ and $TM_{110}$, respectively. Fig. 6 shows the simulated 3D radiation patterns at 20 GHz, 30.3 GHz (RN1), 31.6 GHz (RN2), and 33.2 GHz (RN3) during the stopband. The antenna exhibits full-angle filtering characteristics, with a maximum gain of -19.1 dBi at 20GHz at lower frequency, and a maximum gain of -25.1dBi, -25.9 dBi, and -31.7 dBi at the three radiation nulls at upper frequency at 30.3 GHz, 31.6 GHz and 33.2 GHz respectively.

As shown in Fig. 8(a), to verify the proposed operating mechanism and performance of the Filtenna I, it has been fabricated and measured. Keysight vector network analyzer (VNA) is used to measure reflection coefficients. Radiation patterns are measured in a microwave far-field chamber. The simulated and measured reflection coefficients, and realized gain are given in Fig. 8(b). The simulated and measured -10 dB FBW are 7.1% (27.14 – 29.13 GHz) and 8.6% (27.62 – 30.11 GHz). The simulated and measured realized gain match better and the average gain is above 4.7 dBi in the passband. Fig. 9 gives the simulated and measured normalized radiation patterns at two points, 27.6 GHz and 28.9 GHz. The filtenna patterns are symmetrical both on the E-plane and H-plane. And the consistency at these two frequency points is also high. Besides, the cross-polarized are more than 20 dB for E-plane and H-plane in the broadside direction.

### D. Coupling Topology and Coupling Matrix of Antenna II

In Filtenna II the same substrate thickness as Filtenna I has been used. As it is shown in Fig. 10, the slot radiator in Filtenna II is placed at the opposite of the feeding post. Other structures remain

TABLE I

Comparison Between Previously Reported MmWave Filtennas and Proposed Filtenna

| Ref | FBW (%) | Num. of Layers | Rejection Level (dB) | Ave.Gain (dBi) | Size ($\lambda_0 \times \lambda_0 \times \lambda_0$) | Design Complexity* | Cost ** | Beam scanning filtering capability |
|---|---|---|---|---|---|---|---|---|
| [4] | 29.3 | 5 | >14 | 5.5 | 0.52×0.52×0.12 | High | High | Not available |
| [8] | 12.7 | 3 | >30 | 8.2 | 0.68×1.13×0.2 | High | High | No |
| [12] | 25.8 | 3 | >16 | ≈7.5 | 1.34×1.34×0.172 | High | High | No |
| [15] | 1.56 | 4 | >35 | 6.7 | 0.92×1.06×0.1 | Mid | High | No |
| [17] | 18.4 | 2 | >19 | 10.1 | 1.61×1.43×0.14 | High | Mid | No |
| [18] | 23.9 | 5 | >20 | 4.5 | 0.53×0.53×0.158 | High | High | Yes/±45° |
| [19] | <6.56 | 2 | >24 | <5.08 | 0.7×1.2×0.08 | Mid | Mid | No |
| **This Work** | **8.6** | **1** | **>16** | **4.7** | **0.64×0.64×0.09** | **Low** | **Low** | **Yes/±40°** |
| | **10.1** | **1** | **>23** | **5.3** | **0.64×0.64×0.09** | **Low** | **Low** | **Yes/±40°** |

*Design complexity: Design parameters within 10 are defined as low complexity, between 10 and 20 are defined as medium complexity, and above or equal to 20 are defined as high complexity.

**Cost: one-layer PCB is low cost, two-layer PCB is medium cost, above two-layer PCB is high cost.

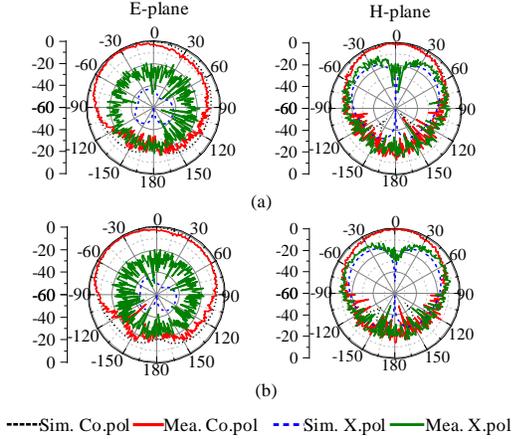

Fig. 14. Simulated and measured normalized radiation patterns of the proposed Antenna II, (a) 28.2 GHz, (b) 29.7 GHz.

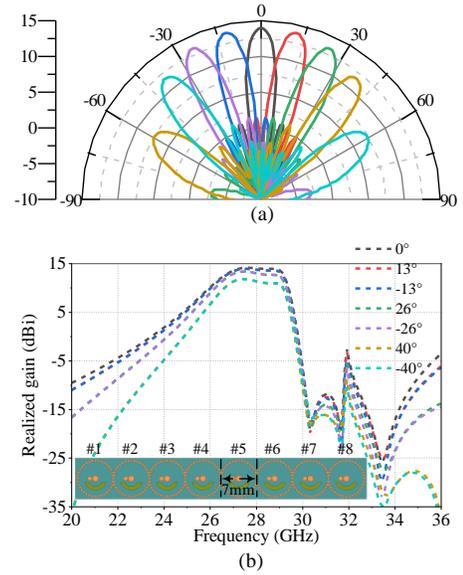

Fig. 15. Simulated beam scanning and realized gain of the proposed Filtenna I. (a) Beam scanning at 28 GHz, and (b) Scanning realized gain.

unchanged, but some structure parameters need to be adjusted. In the Antenna II, there is only one radiation null at the upper band. However, compared with the Filtenna I, its increased low-frequency roll-off, for example, the stopband level at 24 GHz increases from -5dB to -12 dB.

Fig. 11 presents the 3D pattern at 20 GHz and the upper-frequency radiation null of 32.8 GHz. The lower-frequency radiation suppression under this structure increases by nearly 10 dB. At 20GHz, a suppression level of more than 38 dB was obtained at all angles, and a suppression level of more than 27 dB was obtained at the upper-frequency band radiation null. The target design FBW is 6.9% from 28 to 30 GHz with the 12 dB return loss within the passband. Similarly, a reference coupling matrix is given as a matrix:

$$M = \begin{bmatrix} & S & 1 & 2 & L \\ S & 0 & 1.12 & 0.67 & 0 \\ 1 & 1.12 & 1.95 & 0 & 1.12 \\ 2 & 0.67 & 0 & -2.02 & -0.67 \\ L & 0 & 1.12 & -0.67 & 0 \end{bmatrix} \quad (18)$$

The calculated $Q_{ex1}$ and $Q_{ex2}$ are 15.5 and 43.4, respectively. The fabricated prototype of the Antenna II is in Fig. 12. The antenna has a very compact structure. The simulated and measured reflection coefficients, realized gain, and antenna efficiency are shown in Fig. 13. The simulated and measured $|S_{11}|$ are 7.4% (27.86 – 29.99 GHz) and 10.1% (28.11 – 31.09 GHz), respectively. The simulated average in-band gain exceeds 5.3 dB. The in-band gain of Filtenna II is smoother than the Filtenna I. This is because for the $TM_{110}$ mode, the path of the excitation slot is shorter. Although the upper sideband has only one radiation null, its stopband suppression can still exceed 18 dB. The simulation and measurement results are in good agreement. The radiation patterns at 28.2 GHz and 29.7 GHz in these two resonance points are shown in Fig. 14.

### E. Beam Scanning Capability

To meet mmWave transmission requirements in real scenarios, the antenna should provide good beamforming performance [1]. Therefore, two $1 \times 8$ linear antenna arrays for Filtenna I and Filtenna II are simulated to verify the beam scanning performance. The scanning performance at center frequency 28 GHz is provided in Fig. 15 and Fig. 16. Adjacent filtenna elements are connected through a shared row of via in the array (see Fig. 15 (b)). For Filtenna I, the simulated peak scanning gain at 0° is 14.03 dBi and a scanning loss is less than 3 dB at ±40°. The sidelobe levels of the scanning beams are better than 4 dB. As can be seen from Fig. 16(b), the frequency response under each scanning beam has similar filtering characteristics, and the upper and lower sideband suppression can exceed 15 dB. For Antenna II, the beam scanning range it supports is the same as Filtenna I. Each beam of antenna II also has good filtering characteristics and has more than 15 dB stopband suppression in the upper and lower sidebands. Therefore, a good array filtering characteristics can also be achieved using the proposed single layer, and low cost filtenna.

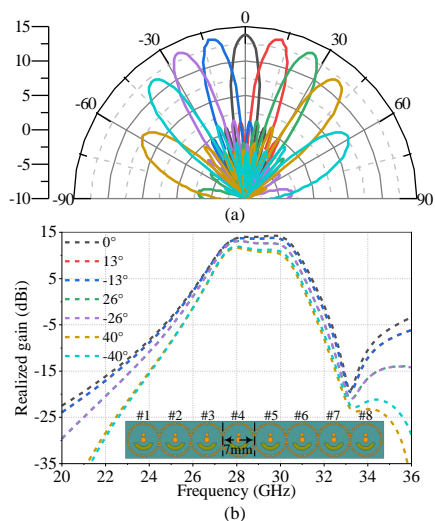

Fig. 16. Simulated beam scanning and realized gain of the proposed Antenna II at 28 GHz. (a) Beam scanning. (b) Scanning realized gain.

### F. Comparison and Discussion

Table I presents the comparison between reported and proposed filtennas with some key performances including bandwidth, rejection level, gain, size, PCB layers, design complexity, costs, and beam scanning capability. The reported filtennas in Table I are all designed with at least 2 layers of PCB. For instance, 5-layers of PCB have been used in [18]. More than one number of layers undoubtedly increases the design cost in practical applications. The antenna element of antennas [12], [15] and [17] occupies larger areas (i.e. >one-wavelength), which indicates that the antenna is only suitable for fixed beam arrays, which greatly limits the practical application of the antenna. Besides, antennas [4], [8], [12], [17], and [18] all use more complex structures to achieve filtering functions, such as the more complex radiation structures in [4] and [12], and the more complex feed structure in [18]. This places higher requirements on the antenna's rapid reproduction, robustness, and processing accuracy. However, the proposed filtenna not only uses single-layer PCB which reduces the cost but also provides a good-filtering response and beam scanning capability which is highly suitable for mmWave applications.

## III. CONCLUSION

Two kinds of novel low-cost filtenna with a low profile are proposed for mmWave applications. A circular FMSIW cavity, a metal post for power feeding, a metal post for improving sideband rejection, and a slot for radiating power is designed to construct the filtenna. In the passband, the slot is excited by the dual modes in the cavity that $TM_{010}$ mode and $TM_{110}$ mode resonate together. It is shown by measurement that the filtenna achieves FBW of 8.6% (27.62 – 30.11 GHz) and 10.1% (28.11 – 31.09 GHz) for these two kinds of antenna. The filtenna features stable radiation patterns with a gain of more than 4.7 dBi. The upper and lower sideband su ppression exceeds 16 dB and 25 dB, respectively. Measured results agree well with the simulated. Therefore, the proposed filtenna can be a good candidate for 5G mmWave indoor applications.